\documentclass[preprint,showpacs,preprintnumbers]{revtex4}
\usepackage{graphicx}
\usepackage{dcolumn}
\usepackage{bm}

\begin{document}

\preprint{APS/123-QED Version 14}
\title{Pressure effect on high-$T_{c}$ superconductors and Casimir Effect in nanometer scale}
\date{\today}
\author{H. Belich, M. T. D. Orlando, E. M. Santos, L. J. Alves, and J. M. Pires}
\affiliation{LETRAF - Phase Transition Studies Laboratory, Departamento de F\'{\i}sica,
Universidade Federal do Esp\'{\i}rito Santo, Av. Fernando Ferrari 514,
Vit\'oria,29075-910 - ES, Brazil }
\author{T. Costa-Soares}
\affiliation{Universidade Federal de Juiz de Fora (UFJF), Col\'{e}gio T\'{e}cnico
Universit\'{a}rio, Av. Bernardo Mascarenhas, 1283, Bairro F\'{a}brica - Juiz
de Fora - MG, 36080-001 - Brasil }
\email{belichjr@gmail.com}

\begin{abstract}
Considering $CuO_{2}$ conducting layers present in high Tc superconductors
as  plasma sheets, we proposed a prescription for the $T_{c}$ pressure
behavior taken into account the Casimir effect. The Casimir energy arises from
these parallel plasma sheets (Cu-O planes) when it take placed in the regime
of nanometer scale (small \textit{d} distance). The charge reservoir layer 
supplies carries to the conducting $nCuO_{2}$ layers, which are a source of 
superconductivity. The pressure induced charge transfer model (PICTM) makes use 
of an intrinsic term, which description is still unclear. Considering Casimir 
energy describing the $T_{c}$ for the case of hight-$T_{c}$ superconductors, 
we propose an explicit expression to the intrinsic term.
Realistic parameters used in the proposed expression have shown an agreement
with experimental intrinsic term data observed in some high-$T_{c}$
compounds.
\end{abstract}

\pacs{74.72.Jt,74.25.Dw,74.62.-c,74.25.Fy}
\maketitle




\section{Introduction}

The high Tc cuprates superconductors have been described since its discovery
in 1986 \cite{Berd} as composed by two major constituents in their unit
cell; $MBa_{2}O_{4-d}$ as a charge reservoir block (M=Cu,Tl,Hg,Bi,C) and a
conducting $nCuO_{2}$ layers (n=2,3,4,5,6). The charge reservoir layer
supplies carries to the conducting $nCuO{2}$ layers and these carriers in $%
CuO_{2}$ planes are a source of superconductivity.

In 1993, Putilin \textit{et al.} \cite{putilin} have obtained a new family $%
HgBa_{2}Ca_{n-1}Cu_{n}O_{y}$ (n=1,2,3 ...), which has presented the highest $%
T_{c}$ (134K) for n=3. This Hg-cuprate system loss its superconducting
properties due to $CO_{2}$ contamination, however this matter has been
overcome by partial substitution of mercury (Hg) by rhenium (Re) \cite{Shi1,Kis}.
 Taken into account the rhenium (Re) substitution, its was
possible for other research groups to study the physical properties of this
family without problems like sample degradation and with a precise oxygen
content control.

Our group has investigated the $Hg_{0.8}Re_{0.2}Ba_{2}Ca_{2}Cu_{3}O_{8+%
\delta}$ in the ceramic form (polycristaline) since 1998 \cite{sin,mtdo}. 
This compound can be described as three $CuO_{2}$ conducting planes
separated by layers of essentially insulating material, which is a feature
that high Tc cuprates have in common.

The mercury family $HgBa_{2}Ca_{n-1}Cu_{n}O_{y}$ (n=1,2,3 ...) has a number
of $CuO_{2}$ conducting layers proportional to $n$. Taken into account the
existence of these $nCuO_{2}$ conducting layers in cuprate superconductors,
in 2003 it was indicated \cite{kempf} that Casimir effect \cite{casimir}
should occur between the parallel superconducting layers in high Tc
superconductors. For ideal conductors layers separated by vacuum the Casimir
energy is described as:

\begin{equation}
E_{c}(d)=-\frac{\pi ^{2}\hbar cA}{720d^{3}}
\end{equation}

where $A$ is the plate's area, and is larger as compared with the distance \textit{d}. 
This equation describes the Casimir energy for two parallel plasma sheets
with larger separation \textit{d}.\newline

Our previous study about $Hg_{0.8}Re_{0.2}Ba_{2}Ca_{2}Cu_{3}O_{8+\delta}$ \cite{mtdo1}, 
with optimally oxygen content ($\delta = 8.79$ and $T_{c}^{max}=133$%
K), has indicated $dT_{c}/dP$=$1.9(2)$K$GPa^{-1}$. Considering that the
optimally oxygen content represent the optimally condition for carrier
transport in the {\it{Cu-O}} cluster formed by $3-CuO_{2}$ layers, we attributed
the $T_{c}$ increment with external pressure as a reduction of the \textit{d}
distance between the {\it{Cu-O}} clusters. In this scenario we propose a investigation 
of the nanometric distances correlation between {\it{Cu-O}} clusters and $T_{c}$ 
variation in the frame of Casimir energy.

\section{Casimir effect in high-$T_{c}$ superconductors}

In high-$T_{c}$ superconductors the hole of the Casimir plates can be attributed
to the $nCuO_{2}$ layers, which form a {\it{Cu-O}} non-superconducting charge carriers 
layers initially, and are able to form the superconductors layers below $T_{c}$. 
As these superconducting {\it{Cu-O}} cluster of layers are separated by two orders of 
magnitude smaller than the London penetration depth, the Casimir effect is reduced 
by several orders of magnitude. Taken into account a small \textit{d} (nanometric scale),
Bordag \cite{bordag} has proposed for a Transverse Magnetic (TM) mode, a modification
on Casimir energy, as following:

\begin{equation}
E_{c}(a)=-5.10^{-3}\hbar cAd^{-5/2}\sqrt{\frac{nq^{2}}{2mc^{2}\epsilon _{0}}}
\end{equation}

In the equation (2) $A$ is the sheet area, $d$ is a nanometric distance between the
sheets, and $n$ represents the carrier density. 

In the regime of small distances (nanometric scale) between the clusters of {\it{Cu-O}} layers, 
the Casimir effect becomes a van der Waals type effect dominated by contributions
from TM surface plasmons propagating along the {\it{ab}} planes \cite{kempf}. Within the
Kempf model, the superconducting condensation energy is the same order of magnitude 
as the Casimir energy. Taken into account the density of states in the case
of a Fermi gas in two dimensions, the transition temperature $T_c$ was
predicted \cite{kempf, kempf1} as below:

\begin{equation}
T_{c}=\frac{2^{3/4}\pi^{1/2}\hbar^{3/2}e^{1/2}n^{1/4}} {10\eta
k_{B}m^{3/4}\epsilon_{0}^{1/4} d^{5/4}}
\end{equation}

The equation (3) presents $m=2*\alpha*m_{e}$ as a carrier effective mass, $n$
as a carrier $CuO_{2}$ layer density, $\eta=1.76$ BSC parameter, $d$ as a 
distance between two $nCuO_{2}$ clusters of layers, $\epsilon_{0}$ as vacuum 
electrical permeability. 

Our interpretation is that $\alpha$ in $m=2*\alpha*m_{e}$ represents a factor
associated with effective mass of the conducting superconductor carrier,
which came from the convolution of local symmetry of $CuO_{2}$ (Ex. Octahedral, 
pyramidal or plane) with the crystal symmetry. The main relation
pointed out by the equation (3) is that $T_{c}$ is a function of $%
\alpha^{3/4}$, $n^{1/4}$, and $d^{-5/4}$. In order to verify the equation
(3), it was built the Table I using realistic values found in our laboratory
and in the literature \cite{mtdo1,Armstrong,Kleche,Novikov,Crisan,Takahashi,
Zhi,Cornelius,Schirber}.

\section{Pressure effect on high-$T_{c}$}

As a high-$T_{c}$ superconductor probe, it was investigated the effect of
hydrostatic pressure under $Hg_{0.82}\-Re_{0.18}\-Ba_{2}\-Ca_{2}\-Cu_{3}\-O_{8+%
\delta}$, labeled here as (Hg,Re)-1223. First of all, to describe the effect of
hydrostatic pressure, it was assumed that the volume compressibility of
(Hg,Re)-1223 is the same one determined for $Hg_{1}\-Ba_{2}\-Ca_{2}\-Cu_{3}%
\-O_{8+\delta}$ compound (labeled as Hg-1223), which is close to 1\%/GPa \cite{hunter}. 
For (Hg,Re)-1223, when the hydrostatic pressure is closer to 0.9GPa, the crystal 
unitary cell volume is reduced down to -0.8\%. The variation of hydrostatic pressure 
up to 1.2 GPa on (Hg,Re)-1223, with different $\delta$ causes different $T_{c}$ 
changes \cite{mtdo1}. The reduction of the unitary cell, under hydrostatic pressure, 
leads to an variation of $T_{c}$ and it is associated to contraction of the 
{\it{a, b and c}}-axis. The different $T_{c}$ dependence, concerning
external hydrostatic pressure, may be interpreted by the pressure
induced charge transfer model (PICTM) modified by Almasan et al. \cite{alm}.
The variation on $T_{c}$ can be described by Neumeier and Zimmermann 
\cite{neu} equation:

\begin{equation}
\frac{dT_{c}}{dP}=\frac{\partial T_{c}^{i}}{\partial P}+\frac{\partial T_{c}%
}{\partial n}\frac{\partial n}{\partial P}
\end{equation}

where the first term is an intrinsic variation of $T_{c}$ with pressure and
the second is related to changes in $T_{c}$ due to variation on the carrier
concentration in $nCuO_{2}$ conducting layers, which are caused by the
pressure's change. For the case of external hydrostatic pressure effects on
samples with optimally oxygen content, such as $%
Hg_{0.8}Re_{0.2}Ba_{2}Ca_{2}Cu_{3}O_{8+\delta}$ with($\delta=8.79$), the
second term in equation (4) vanish. Then, under this condition we have:

\begin{equation}
\frac{dT_{c}}{dP}=\frac{\partial T_{c}^{i}}{\partial P}
\end{equation}

So, for this case, the $T_{c}$ variation will be determined only by the
intrinsic term. The non-negligible intrinsic term $\partial
T_{c}^{i}/\partial P$ suggests an effective contribution of the lattice to
the mechanism of high-$T_{c}$ superconductivity against the role of
carriers. The $Hg_{0.8}Re_{0.2}Ba_{2}Ca_{2}Cu_{3}O_{8.79}$ has shown
 $\partial T_{c}^{i}/\partial P = 1.9 K/GPa$ \cite{mtdo1} and the 
 $YBa_{2}Cu_{3}O{7}$ compound has presented 
 $\partial T_{c}^{i}/\partial P = 0.9 K/GPa$ \cite{neu}.

The intrinsic term has been presented with a physical meaning, but without
an exact description since its introduction in 1992 \cite{alm}. However, if
the $T_{c}$ can be associated to the temperature from Casimir energy, as
proposed by equation (3), the equation (4) can be rewriting as following:

\begin{equation}
\frac{dT_{c}}{dP}=\frac{\partial T_{c}}{\partial d}\frac{\partial d}{%
\partial P}+ \frac{\partial T_{c}}{\partial \alpha}\frac{\partial \alpha}{%
\partial P} + \frac{\partial T_{c}}{\partial n}\frac{\partial n}{\partial P}
\end{equation}

For samples with optimally oxygen content the third term is vanish, as
justified before. For this optimally conditions there is a direct
correspondence between the intrinsic (5) term and the other two significant
terms (6), as specified below:

\begin{equation}
\frac{\partial T_{c}^{i}}{\partial d}= \frac{\partial T_{c}}{\partial d}%
\frac{\partial d}{\partial P}+ \frac{\partial T_{c}}{\partial \alpha}\frac{%
\partial \alpha}{\partial P}
\end{equation}

Substituting the equation (3) in (7) we have an explicit expression to
the intrinsic term, as following:

\begin{equation}
\frac{\partial T_{c}^{i}}{\partial d} = \frac{-5}{4}T_{c}\frac{1}{d}\frac{%
\partial d}{\partial P} + \frac{-3}{4}T_{c}\frac{1}{\alpha}\frac{\partial
\alpha}{\partial P}
\end{equation}

\section{Discussion}

The signal of both terms in the equation (8) are negative, however the
intrinsic term is positive, when the pressure is increase. This behavior can
be justified by the negative signal of the both derivative terms,

\begin{equation}
\frac{\partial d}{\partial P}, \frac{\partial \alpha}{\partial P}
\rightarrow (P\uparrow) \rightarrow \frac{\partial d}{\partial P}<0 ,and ~ 
\frac{\partial \alpha}{\partial P} <0
\end{equation}

The first derivative term represents the compression coefficient in \textit{c%
} axis direction. The crystallography \textit{c} axis is associated with the
nanometric distance \textit{d} between the two $nCuO_{2}$ clusters of layers, located
each one in different adjacent crystals cells. As consequence one can write 
the following expression:

\begin{equation}
K_{c}= - \frac{1}{c}\frac{\partial c}{\partial P}= -\frac{1}{d}\frac{%
\partial d}{\partial P}
\end{equation}

The second derivative term in (9) represents a variation of the effective carrier
mass, which came from the change on the dispersion relation under pressure.
X-ray diffraction and XANES analysis of the (Hg,Re)-1223, with optimally oxygen content
has indicated a tendency of {\it{O-Cu-O}} bond angle being closest $180^{o}$ \cite{luis}. 
The effect of increase the external pressure is to change this {\it{O-Cu-O}} bond angle 
to $180^{o}$. In our point of view, $\alpha$ coefficient is related with the convolution 
of $CuO_{2}$ local symmetry and crystal symmetry. As consequence, $\alpha$ value is 
reduced as comparing with the initial value (ambient pressure), when the pressure is increase. 

\begin{equation}
(P\uparrow) \rightarrow \frac{\partial \alpha}{\partial P}<0
\end{equation}

Therefore, the final signal of $\partial T_{c}^{i}/\partial P$ is positive,
and the intrinsic term will present a positive behavior under external
pressure.

In order to verify the agreement of the equation (8), it was built the Table
II using realistic values found in the literature \cite{hunter,Armstrong,Kleche,Novikov,Crisan,Cornelius,Schirber} 
and in our laboratory for compounds with $nCuO_{2}$ superconducting layers. 
Moreover, the recent discovery (April 2008) of a new superconductor family by 
Takahashi \textit{et al.} \cite{Takahashi} with ($FeAs$) superconducting layers 
suggested that we included the $SmOFeAs$ \cite{Zhi} compound in the Table II also.

The values reduction of effective mass coefficient $\alpha$ suggested a
relation with the total symmetry (local $CuO_{2}$ + crystal) configuration. 
Computing simulation are going on in order to verify the variation of dispersion 
relation in the reciprocal space in order to compare the values suggested in the Table II.

\section{Conclusion}

The Casimir energy was related with the superconducting condensation energy \cite{kempf,
kempf1}, taken into account the density of states in the case of a Fermi gas
in two dimensions. As consequence, the transition temperature $T_c$ was
predicted as function of $m^{3/4}$, $n^{1/4}$, and $d^{-5/4}$. Within this
scenario, the $\alpha$ coefficient in $m=2*\alpha*m_{e}$ was interpreted as 
the effective carrier mass factor from the dispersion relation, taken into 
account the convolution between local symmetry of $CuO_{2}$ (Ex. Octahedral, 
pyramidal or plane) and the crystal symmetry. The values found by $T_{c}$ 
expression is in agreement with the experimental $T_{c}$ values found in the principal
superconductors described in the literature and (Hg,Re)-1223 measured in our
laboratory. The $T_{c}$'s behavior under external hydrostatic pressure
(described by PICTM) shows an intrinsic term, which is identified here 
with the variation of Casimir energy. This intrinsic term's pressure
dependence presents an explicit expression proportional to the
compressibility coefficient of \textit{c} axis and the effective mass of
carrier charge. For the best of our knowledge, the $\partial
T_{c}^{i}/\partial P$ has not presented an explicit expression before. Our
propose describe the dependence of intrinsic term with pressure in agreement
with the values found in the literature.

\begin{acknowledgments}
We would like to thank CNPq Grant CT-Energ 504578/2004-9, CNPq
471536/2004-0, and CAPES for financial supports. Thanks also to Companhia
Sider\'ugica de Tubar\~ao (ArcelorMittal). We gratefully acknowledge to
National Laboratory of Light Synchrotron - LNLS, Brazil (XPD, XAS and DXAS
projects), International Center for Condensed Matter (60 Years of Casimir 
Effect) Brasilia, Brazil, June 2008. H. Belich would like to expresses his 
gratitude to the High Energy Section of the Abdus Salam ICTP, for the kind 
hospitality during the period of this work was done.

\end{acknowledgments}

\newpage


\newpage

\section{List of Tables}

\begin{table}[!h]
\caption{Critical temperature evaluated by Casimir energy $T_{c}^{Cas}$, $%
T^{ref}_{c}$ obtained in the references, and correlations}
\label{temperature}
\begin{center}
\begin{tabular}{lccccccc}
\hline
Compound~ & ~ d(nm) & ~ n($m^{-2}$) & ~ $\alpha$ ~ & ~ $T_{c}^{Cas}$(K) ~ & ~ 
$T_{c}^{ref}$ ~ & ~ Cu-O Sym. ~ & ~ Crystal Sym. ~ \\ \hline
$La_{2}CuO_{4}$\cite{Schirber} & $1.32$ & $1$x$10^{18}$ & $24$ & $38$ & $40$ & 
tilted Octahe. & Tetra. \\ 
$SmOFeAs$ \cite{Zhi} & $0.85$ & $0.6$x$10^{18}$ & $24$ & $58$ & $55
$ & *tilted Pyram. (Fe-As) & Tetra. \\ 
$YBa_{2}Cu_{3}O_{7}$ \cite{Cornelius} & $1.16$ & $2$x$10^{18}$ & $%
12$ & $92$ & $92$ & distorted Pyram. & Ortho. \\ 
$Hg-1201$ \cite{Kleche}\cite{hunter} & $0.95$ & $1.2$x$10^{18}$ & $%
13$ & $96$ & $98$ & Octahe. & Tetra. \\ 
$Hg-1212$ \cite{Kleche}\cite{hunter} & $1.27$ & $2.4$x$10^{18}$ & 
$7$ & $126$ & $127$ & Pyram. & Tetra. \\ 
$Hg-1223$ \cite{Armstrong}\cite{Kleche}\cite{hunter} & $1.58$ & $%
3.1$x$10^{18}$ & $5$ & $135$ & $134$ & Pyram. & Tetra. \\ 
$(Hg,Re)-1223$ \cite{mtdo1} & $1.56$ & $3.2$x$10^{18}$ & $5$ & $%
134$ & $133$ & Pyram. & Tetra. \\ 
$Hg-1234$ \cite{Novikov} & $1.89$ & $4.4$x$10^{18}$ & $4.5$ & $125
$ & $125$ & Pyram. & Tetra. \\ 
$Hg-1245$ \cite{Crisan} & $2.21$ & $5.5$x$10^{18}$ & $4.5$ & $108$
& $108$ & Pyram. & Tetra. \\ \hline
\end{tabular}%
\end{center}
\end{table}
Tetra is Tetragonal, Orth is Orthorhombic, Pyram means Pyramidal, and
Octahe is Octahedral. The {\it{(Fe-As)}} is a new system (2008).

\begin{table}[!h]
\caption{Intrinsic term pressure dependence evaluated by Casimir energy}
\label{intrinsic}
\begin{center}
\begin{tabular}{lcccccc}
\hline
Compound~ & ~ $K_{c}$($10^{-3}GPa^{-1}$) & ~ $\partial T_{c}^{exp}/\partial P
$ & ~$\partial T_{c}^{i}/\partial P$ ~ & ~$\frac{-5}{4}T_{c}\frac{1}{d}\frac{%
\partial d}{\partial P}$ ~ & ~$\frac{-3}{4}T_{c}\frac{1}{\alpha}\frac{%
\partial \alpha}{\partial P}$ ~ & ~$\frac{1}{\alpha}\frac{\partial \alpha}{%
\partial P}$ ($10^{-3}GPa^{-1}$) ~ \\ \hline
$Hg-1201$ \cite{hunter} & $5.8$ & $1.7$ & $1.7$ & $0.7$ & $1.0$ & $-13.8$ \\ 
$Hg-1212$ \cite{hunter} & $6.0$ & $1.7$ & $1.7$ & $0.9$ & $0.8$ & $-8.4$ \\ 
$Hg-1223$ \cite{Armstrong}\cite{hunter} & $5.6$ & $1.7$ & $1.7$ & $0.9$ & $%
0.8$ & $-7.9$ \\ 
$(Hg,Re)-1223$\cite{mtdo1} & $5.6$ & $1.9$ & $1.9$ & $0.8$ & $1.1$ & $-11.0$ \\ 
$Hg-1234$\cite{Novikov} & $5.8$ & $1.2$ & $1.2$ & $0.8$ & $0.4$ & $-4.3$ \\ 
$YBa_{2}Cu_{3}O_{7}$ \cite{Cornelius} & $4.2$ & $0.9$ & $0.9$ & $0.3$ & $0.6$
& $-8.7$ \\ 
$La_{2}CuO_{4}$ \cite{Schirber} & $1.6$ & $2.3$ & $2.3$ & $0.1$ & $2.2$ & $%
-73$ \\ \hline
\end{tabular}%
\end{center}
\end{table}


\end{document}